%% file: main.tex
\documentclass[acmsmall,screen]{acmart}

\usepackage{minted}
\usepackage{multicol}
 \usepackage{makecell}
\usepackage{subfigure}
\usepackage{color, colortbl}
\usepackage{mathalpha}
\usepackage{subcaption}
\usepackage{wrapfig}
\usepackage{tabularray}
\usepackage{savesym}
\savesymbol{bigtimes}
\savesymbol{widebar}
\usepackage{mathabx}
\restoresymbol{MABX}{bigtimes}
\restoresymbol{MABX}{widebar}

\AtBeginDocument{%
  }

\setcopyright{acmlicensed}
\copyrightyear{2025}
\acmYear{2025}
\acmDOI{}
\acmVolume{}
\acmNumber{}
\begin{document}

\title{PrediPrune: Reducing Verification Overhead in Souper with Machine Learning–Driven Pruning}

\newcommand{\THISWORK}{\emph{PrediPrune}}

%
\author{Ange-Thierry Ishimwe}
\orcid{0009-0000-9277-3227}
\authornote{Both authors contributed equally to this research.}
\email{ange-thierry.ishimwe@colorado.edu}
\author{Heewoo Kim}
\email{heewoo.kim@colorado.edu}
\author{Tamara Lehman}
\email{tamara.lehman@colorado.edu }
\author{\\Joseph Izraelevitz}
\email{joseph.izraelevitz@colorado.edu}
\affiliation{
  \institution{University of Colorado Boulder}
  \city{Boulder}
  \state{Colorado}
  \country{USA}
}

\author{Raghuveer Shivakumar}
\authornotemark[1]
\affiliation{%
  \institution{AMD}
  \city{}
  \country{USA}
  }
\email{raghu@raghu.cc}


\renewcommand{\shortauthors}{Ishimwe et al.}

\input{text/000_abstract}


\settopmatter{printacmref=false}
\setcopyright{none}
\renewcommand\footnotetextcopyrightpermission[1]{}
\pagestyle{plain}


\maketitle

\input{text/100_intro}
\input{text/200_background}

\input{text/300_motivation}

\input{text/400_overview}

\input{text/500_design}

\input{text/600_method}

\input{text/700_eval}

\input{text/800_related}
\input{text/900_conclusion}


\bibliographystyle{ACM-Reference-Format}
\bibliography{references}




\end{document}

%% file: text/000_abstract.tex
\begin{abstract}

\emph{Souper} is a powerful enumerative superoptimizer that enhances the runtime performance of programs by optimizing LLVM intermediate representation (IR) code.
However, its verification process, which relies on a computationally expensive SMT solver to validate optimization candidates, must explore a large search space. 
This large search space makes the verification process particularly expensive, increasing the burden to incorporate \emph{Souper} into compilation tools.

We propose {\THISWORK}, a stochastic candidate pruning strategy that effectively reduces the number of invalid candidates passed to the SMT solver. 
By utilizing machine learning techniques to predict the validity of candidates based on features extracted from the code, {\THISWORK} prunes unlikely candidates early, decreasing the verification workload.

When combined with the state-of-the-art 
approach (\emph{Dataflow}), {\THISWORK} decreases compilation time by 51\% compared to the Baseline 
and by 12\% compared to using only \emph{Dataflow}, emphasizing the effectiveness of the combined approach that integrates a purely ML-based method ({\THISWORK}) with a purely non-ML based (\emph{Dataflow}) method.
Additionally, {\THISWORK} offers a flexible interface to trade-off compilation time and optimization opportunities, allowing end users to adjust the balance according to their needs.

\end{abstract}

%% file: text/100_intro.tex
\section{Introduction}
Superoptimization~\cite{massalin1987superoptimizer} seeks to generate the most efficient optimizations that are semantically equivalent to given program fragments, enhancing code quality through a search process rather than relying on predefined transformations. 
Prior work, \emph{Souper}~\cite{sasnauskas2017souper}, is an enumerative superoptimizer  
which operates by first generating optimization candidates --- each potentially more efficient in terms of complexity and execution cost --- by exploring a vast space of possible instruction combinations. 
After candidate generation, a Satisfiability Modulo Theories (SMT) solver~\cite{de2011satisfiability, barrett2021satisfiability} verifies the validity of these candidates, ensuring the correctness and integrity of the program. 

\emph{Souper} has demonstrated its capability to improve code efficiency, reducing the Clang-3.9 binary size by 4.4\% through optimizations that standard LLVM passes cannot achieve \cite{sasnauskas2017souper,mukherjee2020dataflow}. 
Its versatility is further proven by its successful adoption in both the LLVM and Microsoft Visual C++ compilers, making it a valuable tool for a wide range of compiler optimization tasks \cite{sasnauskas2017souper}.
These advantages have made Souper a widely studied topic in recent research~\cite{mukherjee2024hydra,liu2024minotaur,albert2022super,mukherjee2020dataflow, mukherjee2024peephole,taneja2020testing,garba2019saturn,menendez2017alive,cabrera2021crow}.

Despite these advantages, the exhaustive nature of \emph{Souper}'s enumerative synthesis leads to significant challenges, particularly in terms of compilation time. 
The number of candidates generated for a particular code region is exponential, 
 making the process to verify them computationally intensive and time-consuming~\cite{mukherjee2020dataflow}. 
Additionally, the reliance on SMT solvers for verification adds substantial overhead, as SMT solvers are computationally expensive, especially when dealing with a large number of complex instructions.
As a result, based on our characterization of the SPEC CPU 2017 benchmark, the compilation with superoptimization process takes approximately 37 hours, even with all benchmarks compiled in parallel. While this compilation optimization yields a promising 38\% reduction in runtime, the compilation time barrier remains significant.

To address these challenges, efficient pruning is essential to reduce the verification workload for the SMT solver by eliminating unlikely or invalid synthesized candidates before they reach the solver.
Previous efforts have reduced \emph{Souper}'s overhead through deterministic pruning strategies. 
For example, \emph{Dataflow} pruning \cite{mukherjee2020dataflow} reduces the enumerative search space using fast dataflow-based techniques to discard synthesis candidates that contain symbolic constants and uninstantiated instructions. 
While this approach shortens compilation time, it has its own limitations in coverage and adaptability, working best
on easily analyzable code blocks. 

Additionally, the pruning method lacks flexibility in identifying a diverse range of invalid candidates, potentially missing optimization opportunities that a more adaptable, stochastic approach could capture. 
Finally, it primarily works on integer bitvectors and programs without path conditions, limiting its applicability to simpler code blocks.

To address the limitations of prior state-of-the-art, we propose {\THISWORK}, a stochastic candidate pruning strategy for \emph{Souper}.

Our approach efficiently reduces the search space, leading to shorter compilation times while providing broader coverage and being agnostic to the synthesis strategy employed. 
{\THISWORK} leverages machine learning techniques to predict the validity of candidates based on features extracted from the code, allowing for the early elimination of unlikely candidates. 
By employing a Multi-Layer Perceptron (MLP) classifier \cite{kruse2022multi,desai2021anatomization} trained on features extracted from opcode types and the structures of data-flow graphs, we can effectively predict which candidates are unlikely to pass the SMT solver's equivalence checks.
Among the various classifiers in Scikit-learn~\cite{kramer2016scikit, pedregosa2011scikit}, we confirm that an MLP neural network achieves the highest overall accuracy, precision, recall, and F1-score.

To enhance its effectiveness, we integrate {\THISWORK} with \emph{Dataflow}, yielding better optimizations on complex and intricate code blocks and enabling the pruning of a more diverse range of invalid candidates.
\emph{Dataflow}, a deterministic pruning method, first removes a large portion of trivial invalid candidates, while {\THISWORK} further refines the process by eliminating more complex and diverse invalid candidates. 

In addition, {\THISWORK} offers a flexible interface to trade-off between compilation time improvements and runtime gains, making it adaptable to end-users based on their specific needs and objectives. 
By adjusting the classification threshold in the model, users can prioritize either faster compilation times or maximized optimization opportunities.

The evaluation shows that combining {\THISWORK} with the state-of-the-art pruning approach, \emph{Dataflow} pruning, significantly reduces compilation time by 51\% compared to the Baseline (\emph{Souper} without pruning) and by 12\% compared to using \emph{Dataflow} pruning on its own.
This improvement indicates that {\THISWORK} effectively reduces the verification space by removing the invalid optimization candidates from the critical path and compensating for Dataflow's difficulty in handling more complicated code regions.
Notably, {\THISWORK}'s pruning candidates and strategy are generally orthogonal to those addressed by \emph{Dataflow} pruning, making {\THISWORK} a complementary solution. 

Our contributions are as follows:
\begin{itemize}
    \item We design and implement {\THISWORK}, a stochastic, machine learning-based pruning strategy for the \emph{Souper} superoptimizer.
    
    \item We develop an efficient feature extraction method and an MLP classifier to predict and prune unlikely candidates before SMT solver invocation.
    
    \item We introduce a flexible trade-off mechanism that allows users to balance between faster compilation times and maximized optimization potential.
    
    \item We demonstrate that combining {\THISWORK} with \emph{Dataflow} pruning significantly reduces compilation time and verification overhead compared to both the baseline \emph{Souper} and using \emph{Dataflow} pruning methods alone, while maintaining the performance improvements of the generated code.
\end{itemize}

%% file: text/200_background.tex
\section{Background}\label{200_background}

\emph{Souper}~\cite{sasnauskas2017souper} is an enumerative superoptimizer that identifies missed optimization opportunities by analyzing a program's intermediate representation (IR) and generating more efficient code blocks. It operates through two main components: \emph{enumerative synthesis} and a \emph{Satisfiability Modulo Theories (SMT) solver} .

\input{text/210_enumerative}

\input{text/220_smt}
\input{text/230_workflow}

%% file: text/210_enumerative.tex
\subsection{Enumerative synthesis}\label{210_enumerative}

Enumerative synthesis is a method that systematically generates candidate code sequences in search of more efficient alternatives to existing code segments. 
This process involves exploring a large space of possible instruction combinations to find those that are semantically equivalent to the original code but offer improved performance.
In this context, the code being optimized is represented using two key concepts: the \emph{Left-Hand Side} (LHS) and the \emph{Right-Hand Side} (RHS).

The LHS represents the original code segment extracted from the program's IR. 
It includes a root instruction (usually one that returns an integer-typed value) and all the instructions it depends on, forming a subgraph of the program's data flow graph.

The RHS is the candidate code sequence generated by the enumerative synthesizer. 
The RHS candidates are intended to be semantically equivalent to the LHS but potentially more efficient in terms of execution cost or complexity.

For instance, consider an LLVM IR expression: $x$ $\times$ $2$. 
\emph{Souper} identifies this expression as a LHS candidate for optimization.
Through enumerative synthesis, it might generate RHS candidates like $x << 1$, which shifts $x$ left by one bit, effectively multiplying it by two but potentially executing more efficiently on certain architectures.

%% file: text/220_smt.tex
\subsection{SMT solver}\label{220_smt}
An SMT solver plays a crucial role in validating the correctness of the RHS candidates generated during enumerative synthesis. 
The SMT solver checks whether the LHS and each RHS candidate are semantically equivalent—that is, whether they produce the same output for all possible inputs.

The SMT solver operates by attempting to find an input where the outputs of the LHS and RHS differ.
If the solver verifies no such input exists, it deems the query unsatisfiable, confirming that the LHS and RHS are equivalent for all inputs.
If the solver finds an input that produces different outputs for the LHS and RHS, it deems the query satisfiable, indicating that the RHS is not a valid replacement.

For example, when evaluating the candidate $x << 1$ as a potential RHS for the LHS $x$ $\times$ $2$, the SMT solver checks whether there exists any integer value of $x$ for which ($x$ $\times$ $2$) != ($x << 1$) holds true. 
If the solver determines that no such $x$ exists (i.e., the formula is unsatisfiable), it confirms that $x << 1$ is semantically equivalent to $x$ $\times$ $2$ for all integer values of $x$. 
Therefore, $x << 1$ is confirmed as a valid, more efficient RHS.

In program synthesis, SMT solvers are essential for ensuring that any transformations or optimizations preserve the program's intended behavior. 
By leveraging background theories like arithmetic and bit-vectors, SMT solvers provide a formal method to verify semantic equivalence between code segments.

%% file: text/230_workflow.tex
\subsection{Souper Workflow}\label{230_workflow}
\emph{Souper} begins by analyzing the program's IR to extract potential LHS patterns. 
It examines each instruction to identify those that return integer-typed values, designating them as root instructions for LHS candidates. 
It then performs a backwards traversal to include all instructions it depends on, constructing a subgraph that represents the LHS in \emph{Souper} IR.  \emph{Souper}'s IR is a domain-specific IR that resembles a purely functional, control-flow-free subset of LLVM IR. 

With the LHS patterns identified, \emph{Souper} employs enumerative synthesis to generate RHS candidates. 
It systematically combines instructions --- within specified constraints --- to create code sequences that might serve as more efficient replacements for the LHS. 
The synthesizer aims to produce RHS candidates that reduce execution cost or complexity --- potentially using fewer compute cycles --- while at the same time performing the same computation as the LHS.
Depending on the complexity of the LHS, \emph{Souper} may generate anywhere from a few to several hundred RHS candidates.

Each RHS candidate is then passed to the SMT solver for validation.
The solver constructs a logical formula representing the equivalence of the LHS and RHS.
Then it checks for satisfiability to determine if there exists any input where the LHS and RHS produce different outputs.
An unsatisfiable result confirms the candidate as a valid optimization, and a satisfiable result invalidates the candidate.

Once a valid RHS is found, \emph{Souper} further evaluates the complexity of its instructions using a cost function that estimates the number of cycles required for execution, compared to the LHS, to confirm it as more efficient. 
If confirmed, \emph{Souper} replaces the original LHS in the IR with the optimized RHS. This substitution leads to improved performance by enabling more efficient execution of the optimized code.

\emph{Souper}'s approach leverages the exhaustive search capabilities of enumerative synthesis to generate potential optimizations and employs SMT solvers to ensure correctness. 
By systematically generating RHS candidates and rigorously validating them against the LHS, \emph{Souper} enhances program performance while maintaining semantic integrity. 
This process allows for the discovery of non-obvious optimizations that traditional compilers might miss, contributing to more efficient executable programs.

%% file: text/300_motivation.tex
\section{Motivation}\label{300_motivation}

Despite \emph{Souper}'s ability to discover optimizations that traditional compilers might miss, its strategy introduces significant overhead, hindering its integration into existing compiler infrastructures.
This overhead arises primarily from the costly validation of candidates using an SMT solver, which is inherently slow. 
Because the SMT solver is not fast, the broad search space generated during the synthesis process becomes a significant issue, as each candidate requires time-consuming validation.
Consequently, the combination of a slow SMT solver and a massive number of candidates leads to substantial overhead when integrating \emph{Souper} into production-grade compilers.

\input{text/310_enumerative}
\input{text/330_pruning}

%% file: text/310_enumerative.tex
\subsection{Challenges with Enumerative Synthesis}\label{310_enumerative}

\paragraph{Costly validation with SMT Solver}
After generating candidates, \emph{Souper} must ensure that each RHS is semantically equivalent to the LHS. 
The only way to validate this equivalence is by querying an SMT solver for each candidate. 
Since the SMT solver checks whether the outputs of the LHS and RHS are identical for all possible inputs, this process is extremely computationally expensive.
Each solver invocation can consume significant time and computational resources, especially when dealing with complex instructions or a large number of candidates.

\paragraph{Huge Search Space}
Enumerative synthesis systematically generates all possible combinations of instructions up to a certain size or complexity to find semantically equivalent and more efficient code replacements (RHS) for a given code pattern (LHS).
This exhaustive approach results in an enormous number of candidate RHSs—ranging from a few to several hundred per LHS.
The sheer volume of candidates creates a substantial search space that must be explored. 
Processing this vast search space is computationally intensive and time-consuming, further exacerbated by the need to validate each candidate using the computationally expensive SMT solver.

%% file: text/330_pruning.tex
\subsection{Necessity for Candidate Pruning}\label{330_pruning}

\begin{figure}
\center
    \includegraphics[width=.98\textwidth]{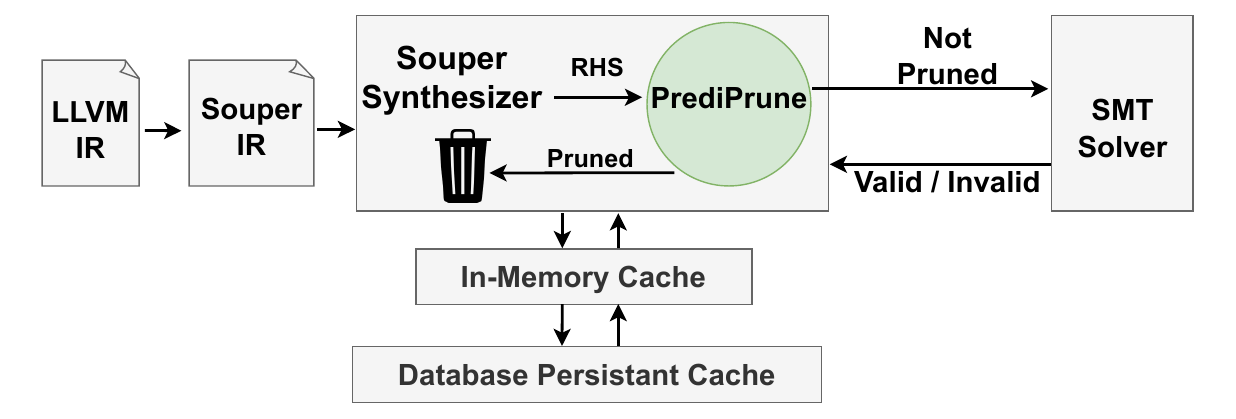}
    \caption{The architecture of \emph{Souper}~\cite{sasnauskas2017souper} enhanced with our proposed {\THISWORK} module, which determines whether a candidate should be sent to the SMT solver or pruned. The cache, when enabled, stores previously validated optimizations.}
    \label{fig:SouperDesign}
\end{figure}

To mitigate these challenges, it is essential to reduce the number of candidates passed to the SMT solver. 
By pruning unlikely or invalid candidates before the expensive validation step, we can significantly reduce the search space, thereby decreasing the compilation time. 
Effective pruning techniques can help make \emph{Souper}'s powerful optimization capabilities more practical for integration into existing compiler infrastructures.
Our work focuses on developing strategies to prune candidates effectively, thereby enabling \emph{Souper} to deliver its optimization benefits without incurring prohibitive overheads.

%% file: text/400_overview.tex
\section{{\THISWORK} Overview}\label{400_overview}
{\THISWORK} introduces a novel and powerful machine-learning-based pruning algorithm for \emph{Souper}, which significantly reduces compilation time.
{\THISWORK} reduces compilation time by decreasing the verification workload for the SMT solver through pruning synthesized candidates using an ML-based model. 

It also offers a flexible interface to trade-off between compilation time improvements and runtime gains, making it possible for end-users to choose the level of optimizations based on their specific needs and objectives. By adjusting the machine learning model parameters, {\THISWORK} can be tuned to prioritize either faster compilation time or maximized optimization opportunities.

The high-level flow of {\THISWORK}~(shown in Figure~\ref{fig:SouperDesign}) begins with \emph{Souper}'s optimization candidates (RHS's), generated through enumerative synthesis. These RHS candidates, along with the LHS (as shown in Figure~\ref{fig:lhsrhs}), are then passed to {\THISWORK} for pruning. During the pruning phase, features of the code are first extracted and subsequently fed into an MLP classifier, which identifies potentially invalid candidates for pruning. The remaining unpruned candidates are forwarded to the SMT solver for further verification.

%% file: text/500_design.tex
\section{{\THISWORK} Design}\label{500_design}

{\THISWORK} has two primary objectives. 
First, it aims to reduce compilation time by pruning as many invalid synthesized candidates as possible before verification. 
Second, it seeks to preserve the effectiveness of \emph{Souper}'s optimizations by ensuring that potentially valid optimizations remain in the pool of candidates.
To accomplish these objectives, \THISWORK{}'s pruning process consists of two key steps: feature extraction and Multi-Layer Perceptron (MLP) classification.

\input{text/510_feature}

\input{text/520_classify}

%% file: text/510_feature.tex
\subsection{Feature Extraction at Multiple Granularities}

The first step of \THISWORK{}'s pruning is extracting features from the LHS and RHS candidates for the classification model.

The features are the result of a combination of techniques that identify properties in the code that are unique for each LHS and RHS candidates. 
We first use dataflow graphs~\cite{appel1998modern}, to capture both the structure and semantics of code blocks. 
For each code block, we extract data from the candidates, such as opcodes, opcode sequences, and computing distances between code blocks, totaling 20 different features (shown in Table \ref{tab:features}).
We use these features to calculate code similarity scores~\cite{ragkhitwetsagul2018comparison} for each RHS candidate compared to its LHS to decide whether to prune the candidate. 

\begin{table}
    \centering
    \resizebox{0.9\textwidth}{!}{%
\begin{tabular}{|l|p{0.6\textwidth}|}
        \hline
        Feature Name & Description 
        \\  \hline 
        1. Inclusion Compression Divergence (ICD)~\cite{ragkhitwetsagul2018comparison} & 
        Compresses the LHS, RHS, and the concatenation of LHS and RHS strings using the Lempel–Ziv–Welch (LZW) compression algorithm, which identifies repeated sequences and replaces them with shorter codes stored in a dynamically built dictionary.  
        It then calculates the difference in size between the compressed concatenation of LHS and RHS strings and the sum of their individually compressed sizes.

        \\  \hline 
        
         2. Longest Common Subsequence (LCS)~\cite{ragkhitwetsagul2018comparison} & Calculates the ratio of the number of characters in the longest common subsequence between the LHS and RHS to the total number of characters in the LHS.
        \\  \hline 
        3. Cosine Similarity~\cite{zakeri2023systematic, ragkhitwetsagul2018comparison}& 
        Using the dataflow graph of block of instructions with operands as nodes, we construct a vector using the number of edges in each node as a ratio of all the nodes, also known as the degree of centrality.
        We generate two vectors from the LHS and RHS, respectively, and calculate the ratio of their inner product to the product of their magnitude.
        \\  \hline 

        4. Number of Constants &  Compares the number of constants used in both sides. 
        \\  \hline 

        5. Number of Arithmetic Operations& Calculates the difference in the number of total arithmetic operations in the LHS and RHS. 
        \\  \hline 
        6. Dice-Sørensen Coefficient~\cite{zakeri2023systematic, ragkhitwetsagul2018comparison} & Calculates twice the ratio of shared opcodes between the LHS and RHS to the sum of their opcodes.
        \\  
        \hline 
        7. Tversky Index~\cite{zakeri2023systematic, ragkhitwetsagul2018comparison} & Calculates the ratio of shared opcodes between the LHS and RHS to the weighted sum of their sizes and differences. We assign a moderate weight ($\alpha$=0.5) to the LHS and a smaller weight to the RHS ($\beta$=0.25).  
        \\  \hline 
         8. Jaccard similarity~\cite{zakeri2023systematic, ragkhitwetsagul2018comparison} & Compares the size of shared opcodes between the LHS and RHS to the total number of opcodes in their union.
         \\  \hline 

         9. Overlap Coefficient Max~\cite{zakeri2023systematic, ragkhitwetsagul2018comparison} & Compares the size of the shared opcodes between the LHS and RHS to the size of the larger set. 
        \\  \hline 

        10. Number of Compare Instructions&  Compares the number of conditional branch instructions in both sides.
        \\  \hline 

        11. Instruction Tree Depth & Compares the tree depth after creating a dataflow graph of both sides.
        \\  \hline 
        12. Total Number of Operands& Calculates the difference in the number of unique operands in the LHS and RHS  (e.g., if the instructions are add r1, r2, r4 and mov r4, r1, the count is 3: r1, r2, and r4).
        \\  \hline

        13. Number of Instructions& Calculates the absolute difference in number of instructions between the LHS and RHS. 
        \\  \hline

        14. Total Number of Unique Opcodes & Calculates the number of distinct types of opcodes used in the LHS and RHS (e.g., if the opcodes are add, sub, add, the count is 2: add and sub, not 3) and then uses the difference between the two.
        \\  \hline 

        15. Number of Variables Declared & Compares the number of variables declared in both sides. 
        \\  \hline

        16. Number of Select Instructions& Compares the number of instruction regions corresponding to C/C++ ternary operations.
        \\  \hline 
        
         17. Overlap Coefficient Min~\cite{zakeri2023systematic, ragkhitwetsagul2018comparison} & Compares the size of the shared opcodes between the LHS and RHS to the size of the smaller set. 
        \\  \hline 

        18. Number of Bit-Width Manipulation Operations& Compares the difference in the number of shift operations in the LHS and RHS. 
        \\  \hline

         19. Number of Block Instructions& Compares the number of instructions in both sides that are used to retain the relationship between Phi nodes. These instructions preserve the values selected by the Phi nodes, which are based on control flow.
        \\  \hline 
        
         20. Number of Phi Instruction& Compares the number of Phi instruction (which selects one block from multiple blocks in the control flow path) in the LHS and RHS. 
        \\  \hline

    \end{tabular}
    %
    }
    \caption{List of features considered in \THISWORK{}. The score utility used to determine the validity of a candidate is shown on Figure~\ref{fig:selectkAutoprune}.}
    \label{tab:features}
\end{table}

 \begin{figure}[htp]
\raggedright
\resizebox{0.17\textwidth}{!}{%
    \begin{minipage}[t]{0.2\textwidth}
\begin{minted}{llvm}
    |\emph{LHS}|

%0 = |block| 2
%1|:|i32 = |var|
%2|:|i32 = |mulnsw| 5:i32, %1
%3|:|i32 = |var|
%4|:|i32 = |addnsw| 4:i32, %3
%5|:|i32 = phi %0, %2, %4
|infer| %5
\end{minted} 
    \end{minipage}
}
   \hspace{13mm}
   \vrule width 0.5pt
   \hspace{1mm}
\resizebox{0.17\textwidth}{!}{%
    \begin{minipage}[t]{0.2\textwidth}
\begin{minted}{llvm}
    |\emph{RHS 1}|

%0|:|i32 = |var|
\end{minted}
    \end{minipage}%
}
\resizebox{0.17\textwidth}{!}{%
    \begin{minipage}[t]{0.2\textwidth}
\begin{minted}{llvm}
    |\emph{RHS 2}|

%0|:|i32 = |var|
%1|:|i32 = |var|
%2|:|i32 = sub %1, %0
%3|:|i32 = add %0, %2
\end{minted}
   \end{minipage}%
}
   \hspace{8mm}
\resizebox{0.17\textwidth}{!}{%
   \begin{minipage}[t]{0.2\textwidth}
\begin{minted}{llvm}
    |\emph{RHS 3}|

%0|:|i32 = |var|
%1|:|i32 = |var|
%2|:|i32 = |var|
%3|:|i32 = |mulnsw| 5:i32, %2
%4|:|i32 = |var|
%5|:|i32 = |fshl| %1, %3, %4
%6|:|i32 = add %0, %5
\end{minted}
   \end{minipage}%
}
   \caption{
   Left: An example of a left-hand side (LHS), which is converted from C to LLVM IR, and then to \emph{Souper} IR.
   Right: Examples of three right-hand side (RHS) candidates generated by \emph{Souper} from the LHS. Pruned candidates will be discarded, while non-pruned candidates will be sent to the SMT solver for semantic equivalence checking.
   }
   \label{fig:lhsrhs}
 \end{figure}

Depending on the approach, the granularity of features can vary by capturing different aspects of the code. 
Instead of relying on a single level, such as lines of code, we employ techniques that analyze entire blocks, individual instructions, and even opcodes to maximize pruning efficiency. 
This multi-granularity analysis allows us to refine the feature extraction process and improve the accuracy of candidate pruning.

We explore token-based (e.g., analyzing opcode sequences), text-based (e.g., representing code sequences as strings), metric-based (e.g., computing distances between code blocks), and graph-based (e.g., analysing data flow between operands across entire instruction blocks) approaches.

\paragraph{Token-based approaches.}
Tokens are strings of characters, such as words or opcodes, and token-based methods have become one of the most popular approaches in code similarity research~\cite{zakeri2023systematic, ragkhitwetsagul2018comparison}.
This popularity is largely due to their robust performance when analyzing large codebases and their effectiveness in detecting code changes~\cite{kustanto2009automatic, juergens2009clonedetective, djuric2013source, saini2016sourcerercc, ragkhitwetsagul2019siamese, rajakumari2019comparison, wu2020scdetector, hung2020cppcd, akram2018droidcc}. 
Token-based approaches convert blocks of code into sequences of tokens, which are then compared with token sequences from other code blocks to identify common subsequences or patterns.

Before performing similarity measurements, we apply filtering to group similar opcodes. 
This process preserves the core semantics of the code, such as the type of operations, while making the comparison more generalizable across different code variations.
For example, opcodes such as \emph{mul} and \emph{shl}, both of which perform multiplication-related operations, are grouped under \emph{mul}. 
Additionally, signed and unsigned operations are generalized by reducing opcodes to their core forms, so that \emph{add}, \emph{addnsw}, and \emph{uadd} are all grouped under \emph{add}. 

We explore several token-based similarity metrics (listed below) to quantify the relationship between the LHS and the RHS of candidate optimizations. 
\begin{itemize}
    \item \emph{Jaccard Similarity}~\cite{zakeri2023systematic, ragkhitwetsagul2018comparison} measures the similarity between two sets by comparing the intersection and union of their elements. 
    The Jaccard score is calculated by this equation:

\[
\text{\( J \)(\( \emph{x} \), \( \emph{y} \))} = \frac{| \emph{x} \cap \emph{y} |}{| \emph{x} \cup \emph{y} |}
\]

Here, \emph{x} represents the opcodes of the RHS, and \emph{y} represents the opcodes of the LHS. 
A high Jaccard score, close to one, suggests a strong similarity.

\item 
\emph{Overlap Coefficient}~\cite{zakeri2023systematic, ragkhitwetsagul2018comparison} measures the proportion of shared elements between two sets relative to the smaller set.

\[
\text{\( O \)(\( \emph{x} \), \( \emph{y} \))} = \frac{|\emph{x} \cap \emph{y}|}{\min(|\emph{x}| , |\emph{y}|)}
\]

An overlap coefficient of 1 indicates that the sets are identical, while a score of 0 shows no commonality. 
This method is useful when focusing on the core elements of the two token sets.

\item 
\emph{Dice-Sørensen coefficient}~\cite{zakeri2023systematic, ragkhitwetsagul2018comparison} is a similarity measure that emphasizes shared elements between two sets by giving more weight to their intersection.

\[
\text{\( DSC \)(\( \emph{x} \), \( \emph{y} \))} = \frac{2|\emph{x} \cap \emph{y}|}{|\emph{x}| + |\emph{y}|}
\]

\item 
\emph{Tversky index}~\cite{zakeri2023systematic, ragkhitwetsagul2018comparison} generalizes the Jaccard and Dice-Sørensen coefficients by allowing for adjustable weighting of the sets’ differences. 
It helps refine similarity measurement by introducing parameters $\alpha$ and $\beta$, which control the importance of asymmetric differences.

\[
\text{\( TI \)(\( \emph{x} \), \( \emph{y} \))} = \frac{|\emph{x} \cap \emph{y}|}{|\emph{x} \cap \emph{y}| + \beta(\alpha a + (1 - \alpha)b)}
\]
Where:
\[
\text{a} = {\min(|\emph{x} \setminus \emph{y}| ,|\emph{y} \setminus \emph{x}|)},
\text{b} = {\max(|\emph{x} \setminus \emph{y}| ,|\emph{y} \setminus \emph{x}|)}
\]

In our implementation, we set $\alpha$ to 0.5 and $\beta$ to 0.25 to balance the trade-off between the Jaccard index and the Dice-Sørensen coefficient.

\end{itemize}

\paragraph{Text-based Approach.} 

In this approach, code sequences are treated as strings, and comparisons are made between two source code regions. 
Text-based methods are effective for identifying duplicate code and detecting similar code, even when there are syntactic or semantic modifications.
We adopt the \emph{Longest Common Subsequence}~(LCS) method, which is one of the most widely used techniques for text-based code comparison ~\cite{ragkhitwetsagul2018comparison,roy2008nicad, whale1990identification, luo2014semantics,wise1992detection}. 
For this approach, we consider the entire code block as the unit of comparison and identify the LCS between the LHS and RHS. 
The similarity is then measured by calculating the ratio of the LCS length to the size of the LHS, providing a metric of how much of the original code is preserved.

\paragraph{Metric-based Approach.} 
Metric-based techniques use mathematical properties, such as distances, to measure differences between two code blocks. 
For this approach, we focus on code block granularity and we apply the distance-based method known as \emph{inclusion compression divergence}~(ICD)~\cite{ragkhitwetsagul2018comparison}, which measures the difference between two code blocks based on compression. The ICD distance is calculated as follows:

\[
\text{\( ICD_Z \)(\( \emph{x} \), \( \emph{y} \))} = \frac{Z(\emph{xy}) - Z(\emph{y})}{Z(\emph{x})}
\]
where \emph{Z(x)}  represents the size of the compressed code block generated by the Lempel–Ziv–Welch (LZW)~\cite{welch1984technique} compression algorithm \emph{Z}, and \emph{Z(xy)} represents the compressed size of the concatenation of \emph{x} and \emph{y}. 

The similarity score ($sim(x,y)$ ) is then derived by subtracting the ICD value from 1, where higher scores indicate greater similarity:

\[
\text{\( sim \)(\( \emph{x} \), \( \emph{y} \))} = 1 -  ICD(\emph{x},\emph{y} )\]

Although metric-based techniques like ICD are sometimes used in code similarity research, they are generally considered less effective than other methods~\cite{ragkhitwetsagul2018comparison}.

\paragraph{Graph-based approach.} 

To capture the structural characteristics of a block of code, we use the entire block of instructions as our granularity level and create a data-flow graph (DFG) for each LHS and its corresponding candidates. 
A DFG represents the flow of data between instructions, where nodes correspond to opcodes and edges represent the flow of data between operands. 

To compare the DFGs, we calculate the degree centrality~\cite{zakeri2023systematic, ragkhitwetsagul2018comparison} of each node for both the RHS and LHS, generating two centrality vectors (A and B). 
We then calculate the \emph{cosine similarity} of these vectors to measure the similarity between the two graphs.

\[
\cos \theta = \frac{\mathbf{A} \cdot \mathbf{B}}{\left\|\mathbf{A}\right\| \left\|\mathbf{B}\right\|}
\]

\paragraph{Other Approaches.} 
We develop a static analyzer to extract further statistics from both the LHS and RHS. 
Using this analyzer, we capture metrics such as the \emph{number of instructions}, \emph{opcode frequency}, and the \emph{count of each opcode type} within the code blocks. 
The opcode types we considered included arithmetic operations, bit-width manipulations, phi nodes, variable declarations, and the number of constant variables. 
Then we calculate the absolute difference between the LHS and each RHS for each of these measurements to assess their similarity.

\begin{figure}
\center
    \includegraphics[width=0.9\textwidth]{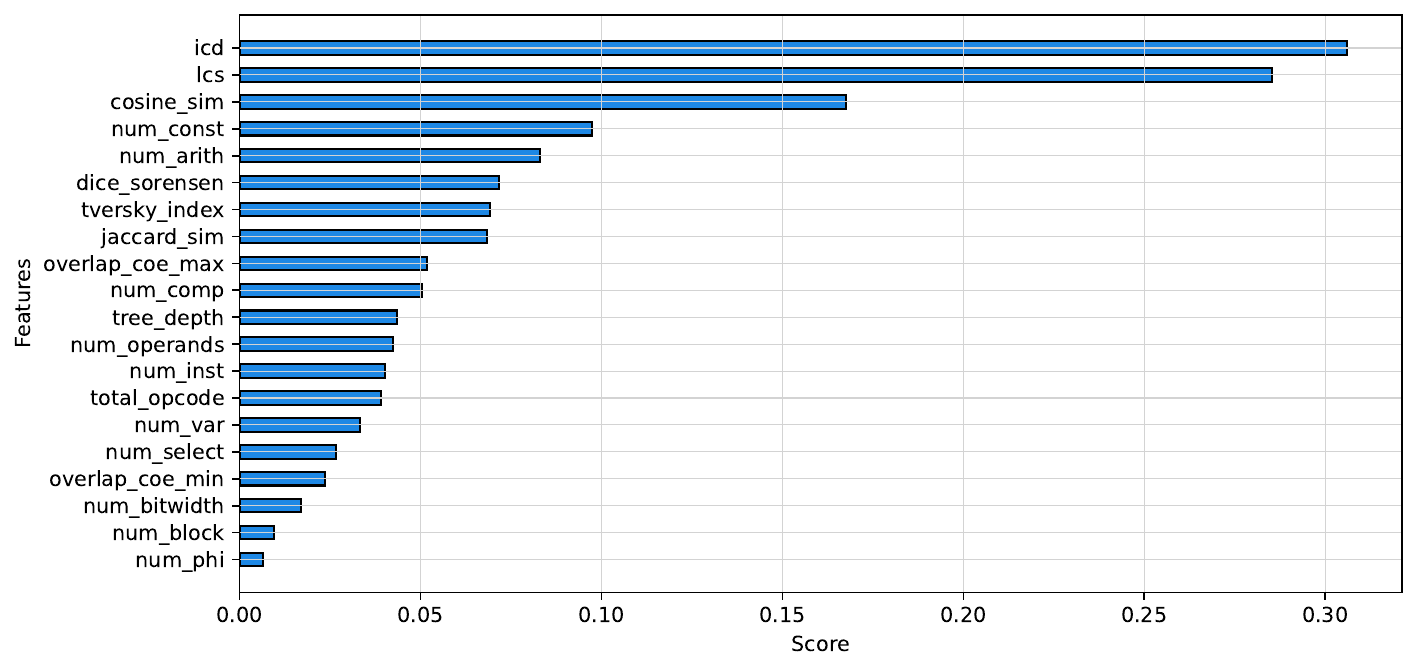}
    \caption{The scores of the features based on their relationship with the output.}
    \label{fig:selectkAutoprune}
\end{figure}

We use Figure~\ref{fig:selectkAutoprune} to demonstrate feature importance, namely, the relationship identified by SelectKBest~\cite{scikitlearnSelectKBest} between each individual feature and its ability to predict whether a RHS is semantically equivalent. In SelectKBest, a zero score indicates that there is no relationship, while higher scores signify a stronger correlation. In Figure~\ref{fig:selectkAutoprune}, the most significant feature is the ICD, with a score of 0.31, whereas the number of phi instructions has a much lower score of 0.007.
Even though ICD shows the strongest correlation, our analysis in Section~\ref{730_features_selection} indicates that feeding the 14 top features to the MLP classifier yields better results than using solely ICD.

%% file: text/520_classify.tex
\subsection{MLP-based Classification of Candidates}\label{520_classify}

The second step of \THISWORK{}'s pruning employs an MLP classifier that identifies likely invalid candidates.
This classifier takes all the similarity scores derived from each extracted feature in the first step as input.
The MLP classifier determines candidates that are likely to be invalid---those with a probability above a certain threshold that the RHS is not semantically equivalent to the LHS. The candidates classified as having a higher probability of being invalid are pruned away from the set of candidates presented to the SMT solver, thereby reducing the search space.
The remaining unpruned candidates are sent to the SMT solver.

\begin{figure}
\center
    \includegraphics[ width=0.6\columnwidth]{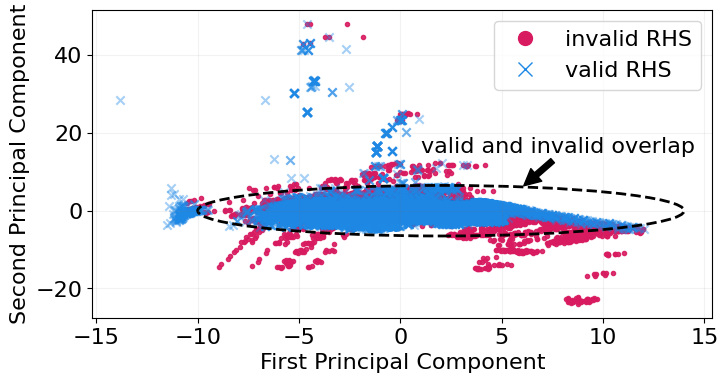}
    \caption{Principal component analysis (PCA) visualization of valid and invalid candidates, indicating that the top two highest scoring features do not capture a significant amount of the available information.  }
    \label{fig:selectK}
\end{figure}

One of the key challenges in pruning candidates is differentiating semantically equivalent code from non-equivalent code without relying on the SMT solver. 
While plagiarism detection and code clone tools often operate on higher-level languages with more contextual information, \emph{Souper} IR lacks this richness, making it harder to differentiate between valid and invalid candidates. 
Although converting \emph{Souper} IR to LLVM IR could provide more information, doing so would introduce significant overhead and negate the efficiency gains of the pruning process.

As illustrated in Figure~\ref{fig:selectK}, using principal component analysis~(PCA) to visualize the features shows considerable overlap between valid and invalid candidates. 
This lack of variance complicates accurate classification, increasing the chances of incorrectly labeling semantically equivalent or non-equivalent RHS candidates.

Given the overlap between valid and invalid candidates, the classification model must carefully balance false positives (invalid candidates misclassified as valid) and false negatives (valid candidates misclassified as invalid).
False positives occur when an invalid RHS is incorrectly classified as valid, increasing compilation time by sending unnecessary candidates to the SMT solver.
False negatives happen when a valid RHS is misclassified as invalid, reducing optimization opportunities and potentially impacting program performance.

Striking the right balance between false positives and false negatives is critical, as it directly influences both compilation time and runtime performance. 
However, this trade-off also provides flexibility for users, allowing them to adjust the classification process based on their specific priorities, whether optimizing for performance or minimizing compilation overhead.  
{\THISWORK} can be tuned to prioritize either faster compilation time or maximized optimization opportunities, as discussed in the following section.

\subsection{Adjusting Optimization Levels: Decision Threshold}

The \textit{decision threshold} of \THISWORK{}'s MLP classifier controls the trade-off between compilation time and the number of optimizations applied. 
By adjusting this threshold, users can influence the classifier's tendency to classify candidate RHS expressions as either valid or invalid. The decision threshold ranges between 0 and 1 and represents the probability cutoff for classifying a candidate as valid. 
If the predicted probability 
meets or exceeds the threshold, it is classified as valid; otherwise, it is classified as invalid.
This flexibility allows users to prioritize what matters most to them—either maximizing performance optimizations or reducing compilation time.

For example, setting a low decision threshold (e.g., 0.1) aims to minimize false negatives—that is, to avoid incorrectly pruning valid candidates—while also allowing more false positives (invalid candidates classified as valid). 
In this case, even though an RHS has a predicted probability of being valid at 0.15, it is considered valid because it exceeds the threshold of 0.1.
An RHS is classified as invalid only when the probability of being valid is less than 0.1 (e.g., 0.05).

Conversely, setting a higher decision threshold (e.g., 0.9) biases the classifier toward minimizing false positives, classifying candidates as valid only if they have a high probability of being correct.
This reduces the number of RHS candidates sent to the SMT solver, shortening the compilation time but potentially limiting optimization opportunities.

Given the trade-offs involved, we provide users the option to adjust the decision threshold based on their specific goals. 
For example, if users prioritize the size and performance of the binary, they may choose a lower threshold (around 0.0001 based on our experimentation), leading {\THISWORK} to classify more candidates as valid to avoid missing optimizations (reducing the benefits of \THISWORK, with only a 12\% speedup). 
Alternatively, if compilation time is the primary concern, a higher threshold (around 0.93 based on our experimentation) can be used, focusing on classifying candidates as invalid unless they have a high probability of being valid RHS, which reduces the solver's workload and speeds up the process (by around 30\% speedup).  
Users can select a decision threshold from a trade-off space derived from the SPEC benchmark measurements (Fig. \ref{fig:paretoFront}) --- these thresholds need not be retrained for each application but can simply be chosen based on our state space exploration.

%% file: text/600_method.tex
\section{Methodology}\label{600_methodology}

\input{text/610_benchmark}

\input{text/620_baseline}

\input{text/630_system}

\input{text/640_training}

\input{text/650_threshold}

%% file: text/610_benchmark.tex
\subsection{Benchmarks and LHS Sampling}\label{610_benchmark}

For the evaluation, we select the SPEC CPU 2017 benchmark suite~\cite{bucek2018spec} because it is widely used to assess general-purpose system performance and offers a diverse set of workloads. 
To focus on assessing {\THISWORK}'s pruning capability, we isolate only the superoptimization phase during compilation after all the \emph{Souper} LHS's are extracted.

We collect all the LHS's from each benchmark and store each one independently in its own file to run optimizations individually. 
This approach is necessary because, although \emph{Souper} is a readily available optimization tool for use with LLVM/Clang, it can be prone to crashes. 
By isolating each LHS, we ensure that if an optimization process crashed, it would be limited to that specific LHS rather than affecting the entire compilation, while still preserving the collected statistics.

After removing duplicate LHS's from each benchmark, we extract a total of 362,495 unique LHS across all workloads. 
However, the number of LHS varies significantly between benchmarks—for example, the \emph{lbm} benchmark has 799 LHS, while \emph{gcc} has 120,460 LHS. 
To ensure a consistent and manageable evaluation across all benchmarks, we run our experiments on a sample of a maximum of 2,000 LHS for each benchmark.
These selected LHS (up to 2,000) are randomly chosen from the pool of unique LHS extracted for each benchmark.
Each of the LHS is distinct, ensuring there are no duplicates.
This sampling strategy allows us to compare the pruning capabilities of {\THISWORK} across different benchmarks under similar conditions, ensuring that the results are not biased by the varying sizes of the LHS pools. Only three benchmarks in the SPEC CPU 2017 suite have less than 2,000 LHS and the smallest one has 747 LHS, so the variation across the benchmarks is not significant.

%% file: text/620_baseline.tex
\subsection{Compilation Strategies}\label{620_baseline}

To evaluate the pruning method, we define several compilation strategies for comparison. 
Our main result is the combined
{\THISWORK} pruning with \emph{Dataflow}.
We also run {\THISWORK} only to isolate its
impact.  We compare
against using only \emph{Souper} (as the Baseline) and the state-of-the-art \emph{Dataflow} pruning method.
\begin{itemize}
    \item \textbf{Baseline.} 
    The \emph{Baseline} implementation is the original \emph{Souper}~\cite{sasnauskas2017souper} without any pruning. 
    We use this as a fixed reference point to assess the impact of various pruning techniques.
    
    \item \textbf{Dataflow.} 
    The \emph{Dataflow} pruning method, as described by Mukherjee et al.~\cite{mukherjee2020dataflow}, is the current state-of-the-art technique used in \emph{Souper}. 
    It employs bidirectional dataflow analysis to eliminate RHS candidates containing symbolic constants or uninstantiated instructions, significantly reducing the search space.
    
    \item \textbf{PrediPrune.}
    The {\THISWORK} approach, in isolation without Dataflow, applies our feature extraction and MLP classification-based pruning method to filter out RHS candidates before querying the SMT solver.
    
    \item \textbf{PrediPrune $+$ Dataflow (main result).} 
    Our combined approach combines the strengths of both \emph{Dataflow} and {\THISWORK}. 
    Initially, the \emph{Dataflow} pruning method is used to remove a large portion of trivial invalid candidates. 
    Then, {\THISWORK} is applied to further prune the remaining candidates. 
    Since the \emph{Dataflow} and {\THISWORK} methods are orthogonal, they can be effectively applied together to reduce the verification workload.
    
\end{itemize}

%% file: text/630_system.tex
\subsection{Experiment Platform and System Specification}\label{630_system}

The experiment is conducted on an ARM AArch64 processor with 32 cores, 1 thread per core, running at a maximum clock frequency of 2.91 $GHz$, and equipped with 125 $GB$ of DDR4 DRAM at 2667 $MT/s$. 
We run all workloads for each pruning method in parallel, but since each LHS is processed independently, they are handled sequentially, one after the other.

A time limit of 300 seconds is set for each individual LHS to conduct RHS candidate generation and verification, with a 5-second timeout for the solver per RHS candidate which corresponds to the timeouts used in \emph{dataflow}'s~\cite{mukherjee2020dataflow}. 
For the RHS candidates generated from each LHS, the candidates are fed into the solver in ascending order of the cost function -- estimating the number of cycles required for execution -- ensuring that the highest-performing candidates are tested first within the 300-second time limit.
All other parameters remain constant across experiments, including the use of the Z3 solver~\cite{de2008z3} and a Redis server~\cite{carlson2013redis} for external caching. 
To compare fairly to Souper and Dataflow we use the Z3 solver in sequential mode as this is the mode they support --- we expect parallelizing the Z3 solver would result in linear speed-ups across all systems and would not change relative results. 

We conducted two experiments to simulate real-world usage of \emph{Souper} with and without an external cache, as illustrated in Figure ~\ref{fig:SouperDesign}.
In the first experiment, we run \emph{Souper} \textbf{without the external cache}, emulating a cold cache scenario typical of the first compilation of a new program. 
Without cached data, \emph{Souper} must generate and validate all optimization candidates from scratch, leading to longer compilation times due to extensive SMT solver validations.
In the second experiment, we enabled the \textbf{external cache}, representing a warm cache scenario common in subsequent compilations or when compiling programs with shared code patterns. 
\emph{Souper} can retrieve previously validated optimizations from the cache, reducing the need for expensive SMT solver invocations and significantly decreasing compilation time.
By comparing these two configurations—cold cache (without external cache) and warm cache (with external cache)—we demonstrate the practical benefits of using \emph{Souper} with caching in real-world applications.

%% file: text/640_training.tex
\subsection{Features Selection}\label{730_features_selection}

\begin{figure}
\center
    \includegraphics[width=0.95\textwidth]{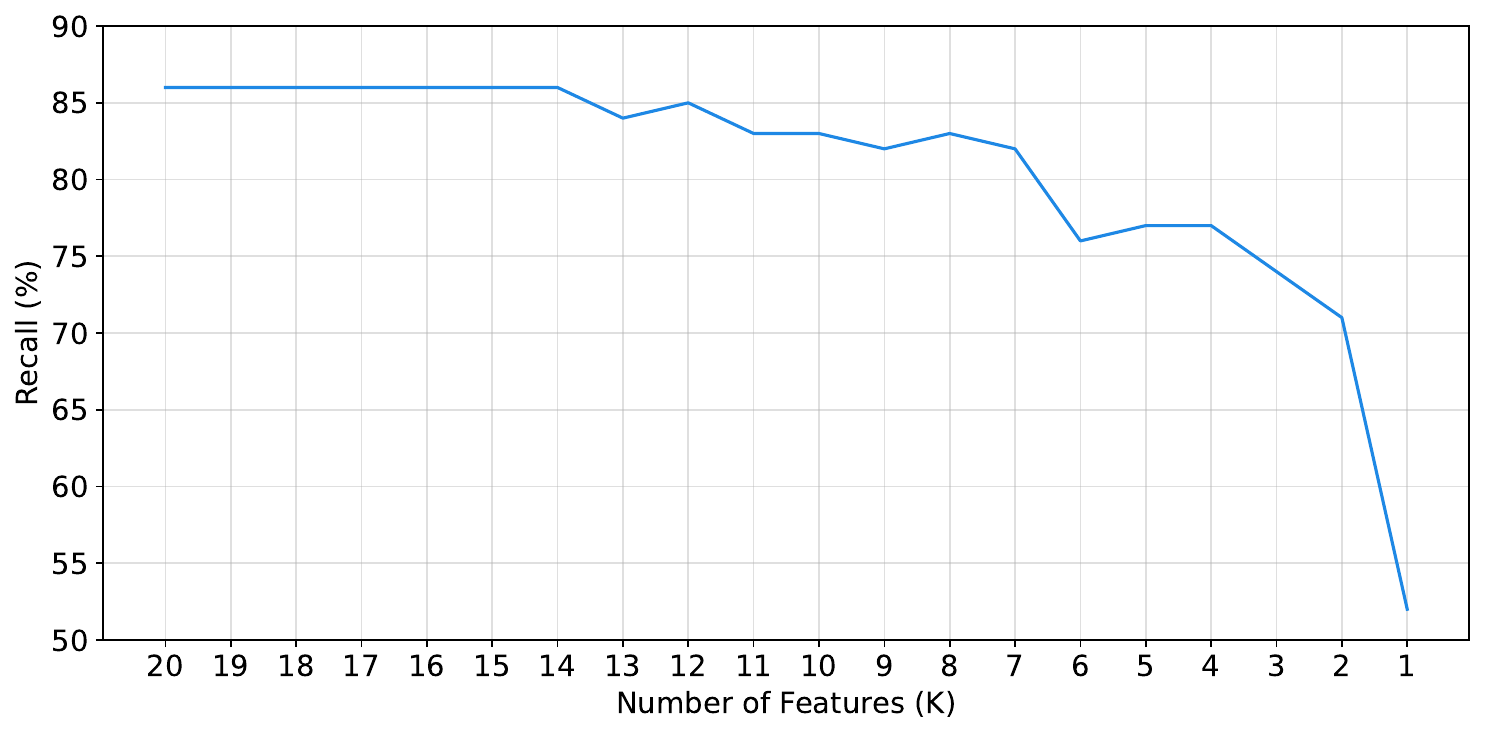}
    \caption{The recall scores for valid candidates while varying the number of features required}
    \label{fig:sensitivity_features}
\end{figure}

While extracting a large number of features is generally advantageous for modeling tasks, their overabundance in machine learning environments can negatively impact performance. 
To address this challenge, we use \emph{SelectKBest} (see Section~\ref{520_classify}), a feature selection algorithm that identifies the top \emph{k} most informative features based on a scoring function. 
We test values of \emph{k} from 20 to 1 with \emph{mutual\_info\_classif} as our scoring metric (shown in Figure~\ref{fig:selectkAutoprune}). We progressively discard features up to the \emph{k} threshold which have the lowest scores during model training.
Figure~\ref{fig:sensitivity_features} highlights the experimental results using the recall percentage: at \emph{k} = 20 (using all features), the model reaches 86\% valid recall with 84\% accuracy, whereas at \emph{k} = 1, valid recall drops to 52\% with 80\% accuracy. The optimal value is \emph{k} = 14 which maintains the maximum recall score of 86\% with an accuracy of 85\%; further reductions in \emph{k} results in declining recall, as shown in Figure~\ref{fig:sensitivity_features}. 
The scores of the discarded features, shown in  Figure~\ref{fig:selectkAutoprune}, reveal that they primarily provide quantitative information. 
For instance, the feature \emph{num phi} (difference in the number of of phi instruction; represented by the 20th bar in Figure~\ref{fig:selectkAutoprune}) has a cumulative total of $\approx$33K instances where this feature is non-zero.
In contrast, \emph{num arith} (difference in the number of arithmetic operations between the LHS and RHS; represented by the 5th bar in Figure~\ref{fig:selectkAutoprune}) has a non-zero value for $\approx$300K instances.
This suggests that the discarded features do not provide sufficient informational value for training.

\subsection{MLP Classification Training Dataset}\label{640_training}

\begin{table}[h!]
    \centering
    \caption{Performance Metrics for Different Models during testing}
    \begin{tabular}{|l|c|c|c|c|}
        \hline
        \textbf{Model} & \textbf{Precision} & \textbf{Recall} & \textbf{F1-Score} & \textbf{Accuracy} \\ \hline
        Multi-layer Perceptron & 0.67 & 0.87 & 0.71 & 0.86 \\ \hline
        Random Forest  & 0.61 & 0.75 & 0.63 & 0.81 \\ \hline
        Logistic Regression & 0.56 & 0.69 & 0.51 & 0.64 \\ \hline
        Naive Bayes    & 0.59 & 0.53 & 0.43 & 0.53 \\ \hline
    \end{tabular}
    \label{tab:model_metrics}
\end{table}

To enable accurate predictions for pruning candidates, we need to create a large and representative dataset for training the MLP classifier. 
We extract left-hand sides (LHS) from several benchmark suites, including GAP~\cite{beamer2015gap}, Coremark~\cite{gal2012exploring}, MachSuite~\cite{reagen2014machsuite}, and MiBench~\cite{990739}. 
Using these benchmarks, we employ \emph{Souper} to generate all possible right-hand side (RHS) candidates for each LHS, capping the number at 300 candidates per LHS to maintain computational feasibility.
The complexity of each LHS and its generated RHS candidates can be evaluated using a cost function, as mentioned in Section \ref{230_workflow}.
Since training is performed on a large number of diverse benchmarks, users do not need to retrain the model for new programs but can simply adopt the provided trained parameters.
Additionally, the compact set of unique instructions in Souper IR, which comprises only 51 integer instructions derived from LLVM IR, enables the model to generalize well across different programs, further reducing the need for retraining~\cite{sasnauskas2017souper}.

For each candidate, we use the Z3 SMT solver to determine its validity—whether the RHS is semantically equivalent to the LHS (valid) or not (invalid). 
This process results in a total of 641,758 candidates generated from 6,534 unique LHS. 
Among these candidates, only 53,377 are valid, representing approximately 8.3\% of the dataset. 
This significant imbalance between valid and invalid candidates poses a challenge for effectively training the classifier, as machine learning models can be biased toward the majority class.

To address this imbalance, we apply a clustering-based under-sampling technique known as ClusterCentroids~\cite{lin2017clustering}. 
This method reduces the number of samples from the majority class (invalid candidates) by clustering them and selecting centroids, thereby creating a more balanced dataset without losing critical information. 
After applying this technique, we obtain a balanced dataset consisting of 85,566 candidates, evenly split between valid and invalid RHS samples with 42,783 instances each.
  
We partition this balanced dataset using an 80/20 split for training and testing purposes. 
For the classification model, we select an MLP neural network implemented in Scikit-learn~\cite{kramer2016scikit, pedregosa2011scikit} as it achieves the highest overall scores in accuracy, precision, recall, and F1-score compared to other classifiers from Scikit-learn as shown in Table~\ref{tab:model_metrics}. 
In addition, we perform a sensitivity analysis to fine-tune the number of hidden layers. 
To determine the best model, we use a cost function that quantifies the execution cost (in cycles) of a block of instructions. This metric enables direct comparison of the performance gains achieved through optimization. As shown in Figure~\ref{fig:NN_sizes}, the MLP model with three hidden layers configured as 16, 32, and 16 neurons achieves the highest score of 50 compared to the alternatives. The remainder of the work is evaluated using this configuration.
We keep our design exploration of machine learning models constrained to models that require minimal amount of resources (computing and memory), to reduce the additional needs of \THISWORK{}. 
We use the \emph{tanh} activation function  for non-linear transformations (see Table~\ref{tab:activation_metrics} for results) and the Adam optimizer for efficient gradient-based optimization (see Table~\ref{tab:solver_metrics} for a comparison with other optimizers). The learning rate is set to 0.01 (see fig~\ref{fig:learning_rate}), as it achieves the lowest loss. Finally,
the network is trained for 400 iterations to ensure convergence and to avoid over-fitting.

%% file: text/650_threshold.tex
\begin{figure}
\center
    \includegraphics[width=0.9\textwidth]{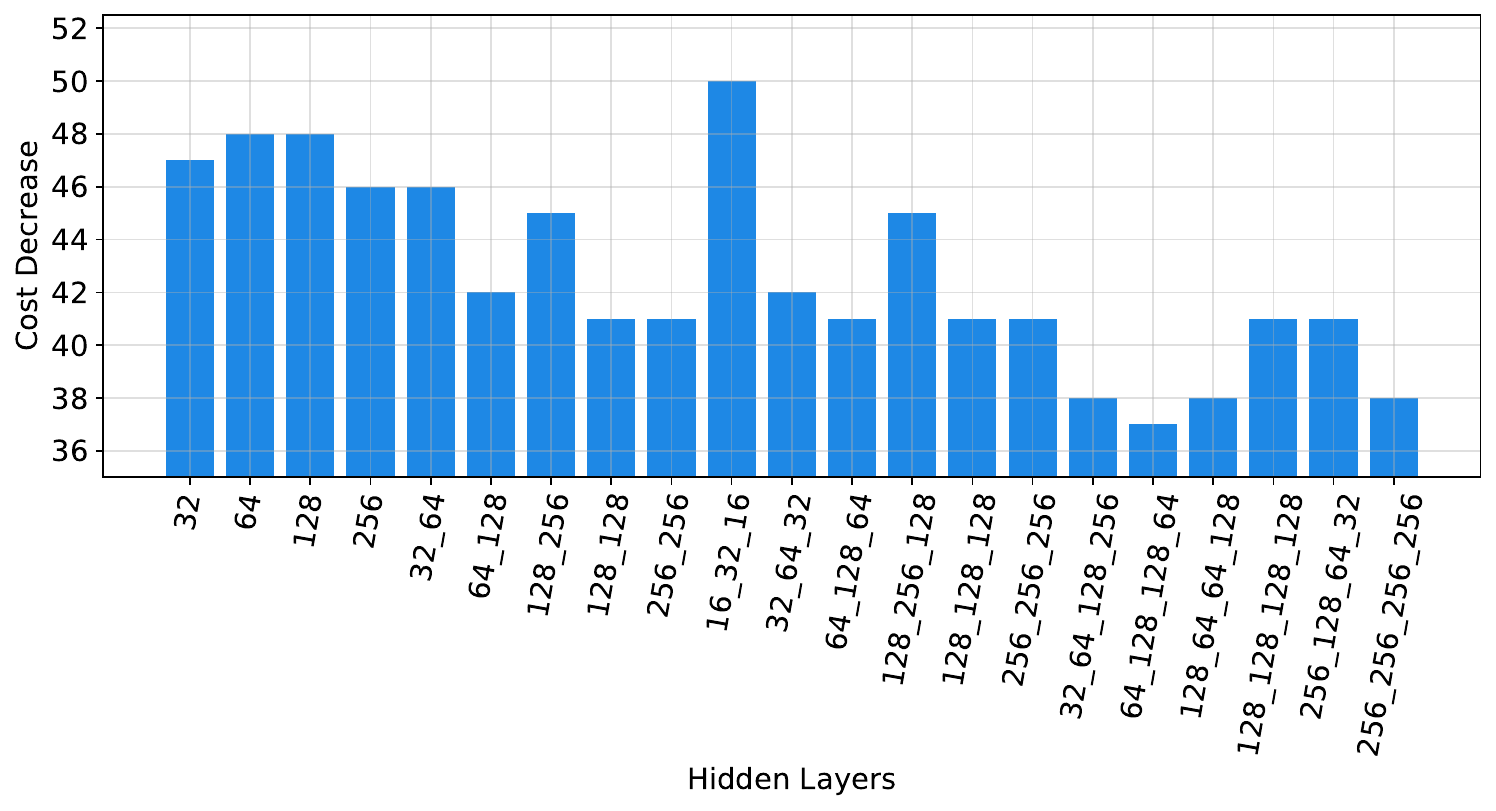}
    \caption{The hidden layers used in the determining the optimal MLP model.}
    \label{fig:NN_sizes}
\end{figure}

\begin{table}[h!]
    \centering
    \begin{tabular}{|l|c|c|c|c|}
        \hline
        \textbf{Model} & \textbf{Precision} & \textbf{Recall} & \textbf{F1-Score} & \textbf{Accuracy} \\ \hline
        relu & 0.66 & 0.86 & 0.70 & 0.85 \\ \hline
        identity  & 0.56 & 0.71 & 0.52 & 0.66 \\ \hline
        logistic & 0.61 & 0.81 & 0.62 & 0.78 \\ \hline
        tanh & 0.67 & 0.87 & 0.71 & 0.86 \\ \hline
    \end{tabular}
    \vspace{0.1in}
    \caption{Performance Metrics for MLP activation functions}
    \label{tab:activation_metrics}
\end{table}

\begin{table}[h!]
    \centering
    \begin{tabular}{|l|c|c|c|c|}
        \hline
        \textbf{Model} & \textbf{Precision} & \textbf{Recall} & \textbf{F1-Score} & \textbf{Accuracy} \\ \hline
        adam & 0.67 & 0.87 & 0.71 & 0.86 \\ \hline
        *lbfgs  & 0.68 & 0.87 & 0.72 & 0.87 \\ \hline
        sgd & 0.63 & 0.82 & 0.65 & 0.80 \\ \hline
    \end{tabular}
    \vspace{0.1in}
    \caption{Performance Metrics for MLP optimizers. \emph{lbfgs} does not converge even when the maximum iterations is set to 10K, and at the risk of overfitting or using an uncoverged model we use \emph{adam} as our optimizer.}
    \label{tab:solver_metrics}
\end{table}

\begin{figure}
\center
    \includegraphics[width=0.9\textwidth]{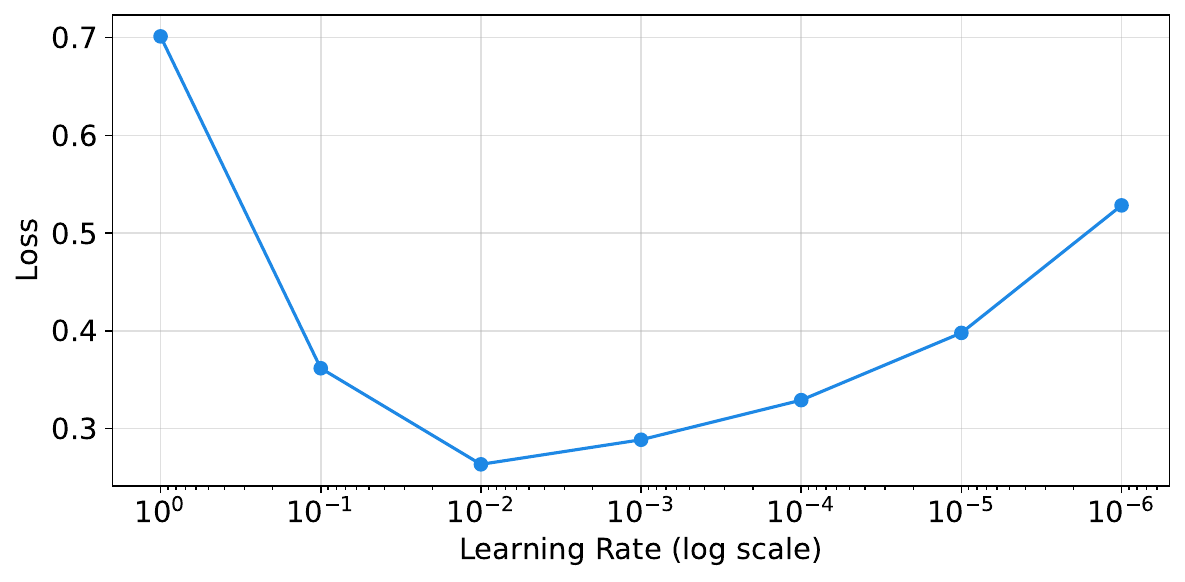}
    \caption{Final training loss of the MLP model at different learning rates.}
    \label{fig:learning_rate}
\end{figure}

\subsection{Decision Threshold Adjustment}\label{650_threshold}

\begin{figure}
\center
    \includegraphics[width=0.9\textwidth]{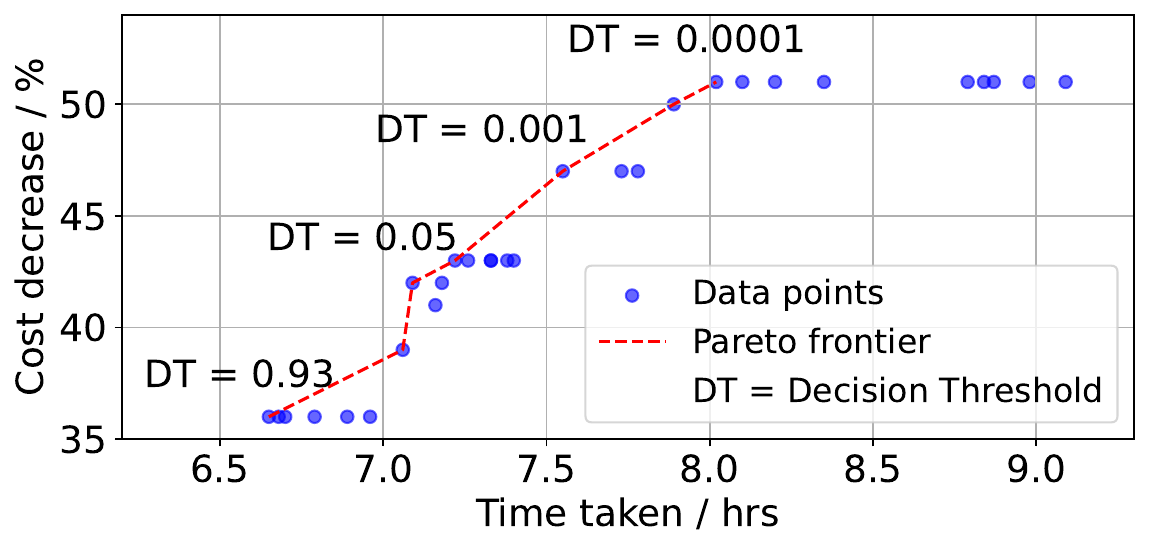}
    \caption{Pareto front used to determine the optimal decision threshold for the PrediPrune $+$ Dataflow implementation.}
    \label{fig:paretoFront}
\end{figure}

\begin{figure}
\center
    \includegraphics[width=0.6\textwidth]{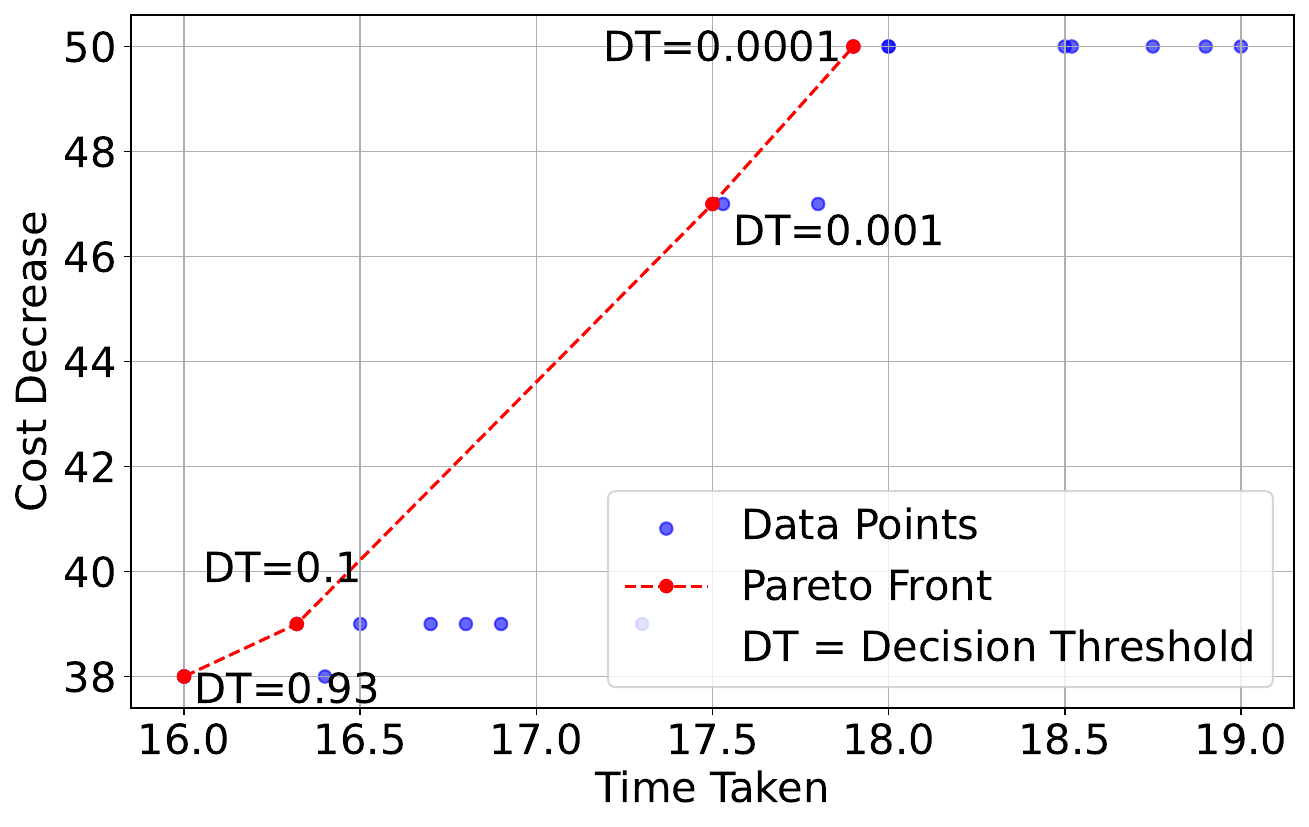}
    \caption{ Pareto front used to determine the optimal decision threshold for the isolated PrediPrune implementation.}
    \label{fig:paretoFrontAutoprune}
\end{figure}

\emph{Souper} provides a \emph{cost function} that measures the execution cost of a block of instructions in terms of the number of cycles required. 
For example, the \emph{add} instruction has a cost of 1 cycle, while \emph{sdiv} has a cost of 3 cycles.  These costs are essential for comparing candidate optimizations --- the lowest cost valid optimization should be used.  Given these costs, we can determine the overall optimization efficiency for a given benchmark by looking at the reduced cost from the original LHS to the valid RHS.

\begin{figure}
\center
    \includegraphics[width=0.49\columnwidth]{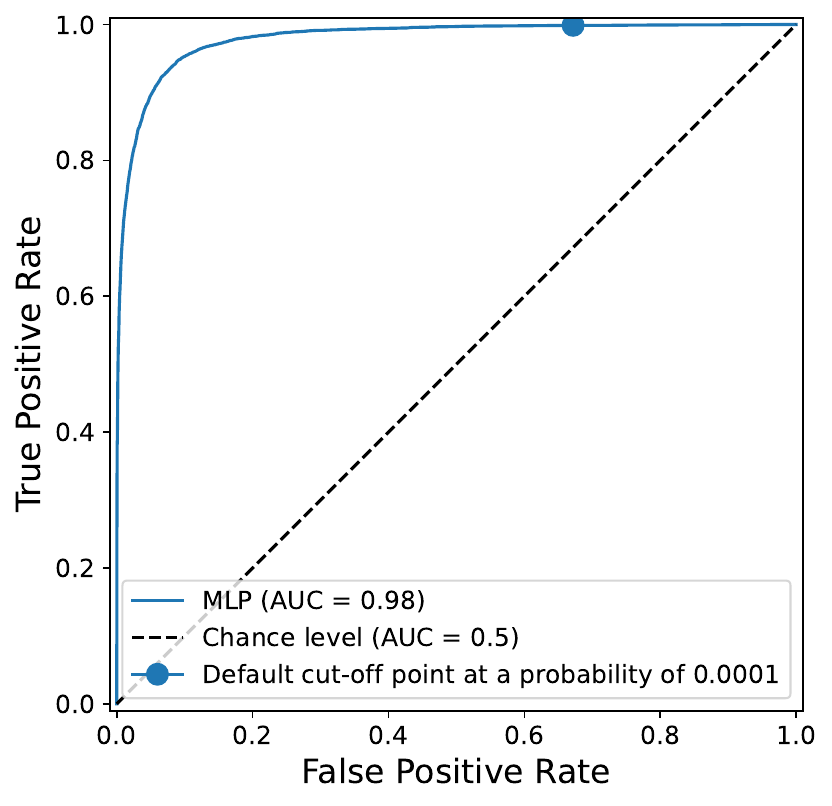}
    \caption{Receiver operating characteristic (ROC) curve for a decision threshold of 0.0001.}
    \label{fig:roc}
\end{figure}

To find the decision threshold that maximizes cost reduction and minimizes compilation time, we plot the pareto front of the possible decision thresholds. 
We choose a representative  benchmark of the geometric mean for the SPEC CPU 2017 benchmark suite, \emph{namd}, based on its Souper behavior (number of successful optimizations, number of failed optimizations, etc).  
The decision thresholds that maximize the cost decrease for both implementation are shown in Figure~\ref{fig:paretoFront}~and~\ref{fig:paretoFrontAutoprune} which are 
determined to be 0.0001
---beyond this point, reducing the decision threshold further only increases compilation time, without corresponding performance improvement. This threshold results in a cost decrease of 52\% when all LHS of \emph{namd} are compared against valid RHS candidates. 
This threshold means that if a candidate RHS has a probability greater than 0.0001 of being valid, it will not be pruned, even if the probability of being invalid is higher.

By setting such a low threshold, we aim to eliminate as many false negatives as possible (i.e., valid candidates incorrectly pruned), accepting an increase in false positives (i.e., invalid candidates not pruned). 
This approach maximizes the retention of valid optimizations at the expense of processing more candidates through the SMT solver.

The receiver-operating characteristic (ROC) curve for the resulting model is shown in Figure~\ref{fig:roc}.
The ROC illustrates the trade-off between the true positive rate and false positive rate at different decision thresholds. 
At a decision threshold of 0.0001, the true positive rate approaches 1, indicating that nearly all valid candidates are retained. 
However, the false positive rate is around 0.6, meaning that some invalid candidates are also classified as valid.
A high recall rate of 99\% and moderate precision of 10\% at this decision threshold also demonstrate the effectiveness of our strategy. 
These outcomes align with our goal of minimizing false negatives to preserve valid candidates, even at the cost of including some false positives.

%% file: text/700_eval.tex
\section{Evaluation}\label{700_eval}

In Figures \ref{fig:evaluation} and \ref{fig:evaluationCache}, we show the compilation results for benchmarks from the SPEC CPU 2017 suite. Our results show that \emph{PrediPrune+Dataflow reduces}  
compilation time by more than 50\% compared to the Baseline and 12\% compared to Dataflow, while maintaining the same level of cost reduction.
As stated, for each benchmark, we sample 2000 LHS and track both the compilation time and the overall optimization achieved, measured using \emph{Souper}'s instruction cost metrics.

\input{text/710_evaluation}

\input{text/720_evaluation_cache}

%% file: text/710_evaluation.tex
\subsection{Performance without an External Cache}\label{710_evaluation}

\begin{figure}[htp]
    \centering
    \begin{subfigure}
        \centering
        \includegraphics[width=0.9\textwidth]
        {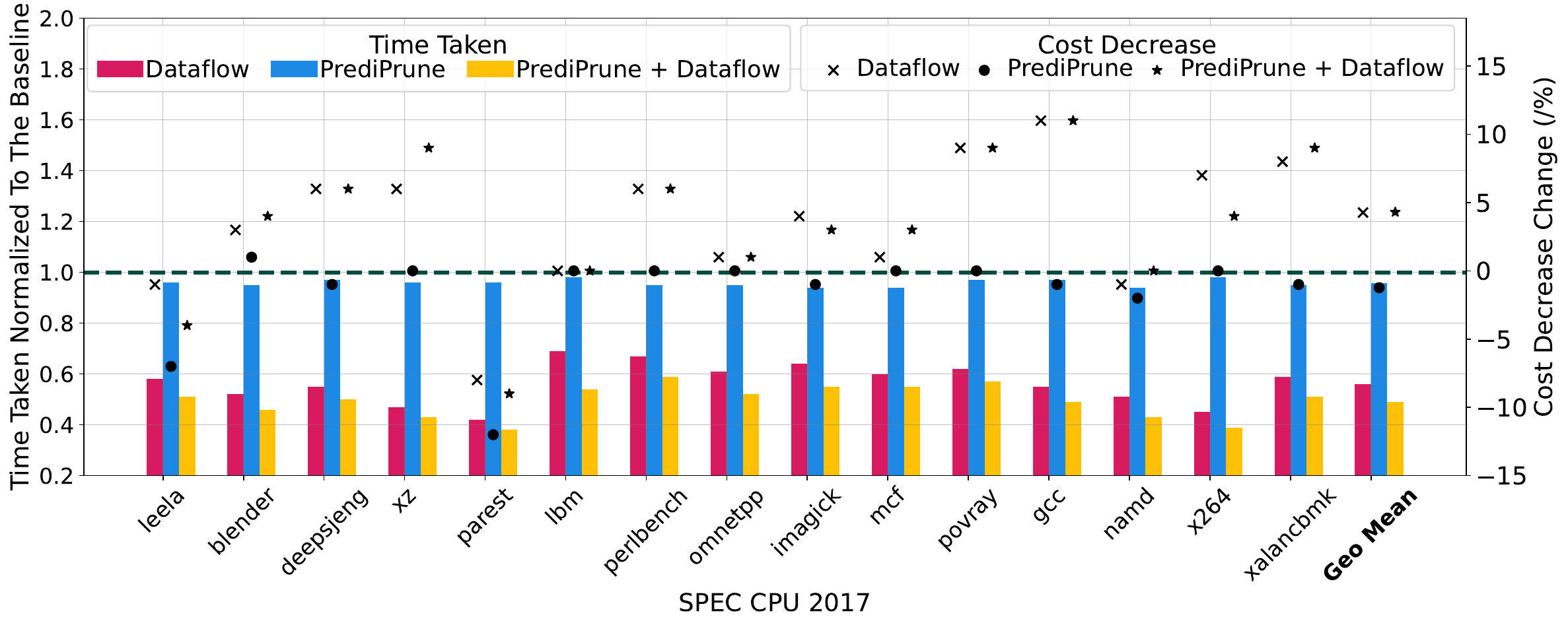}
        \caption{Results of running the SPEC C and C++ SPEC CPU workloads without the external cache configuration.
        The bar chart (left y-axis) illustrates the time taken for optimizations, normalized to \textit{Souper}. The dots (right y-axis) indicate the additional cost decrease achieved relative to \textit{Souper}.}
        \label{fig:evaluation}
    \end{subfigure}
    \vspace{0.5cm}
    \begin{subfigure}
        \centering
        \includegraphics[width=0.9\textwidth]
        {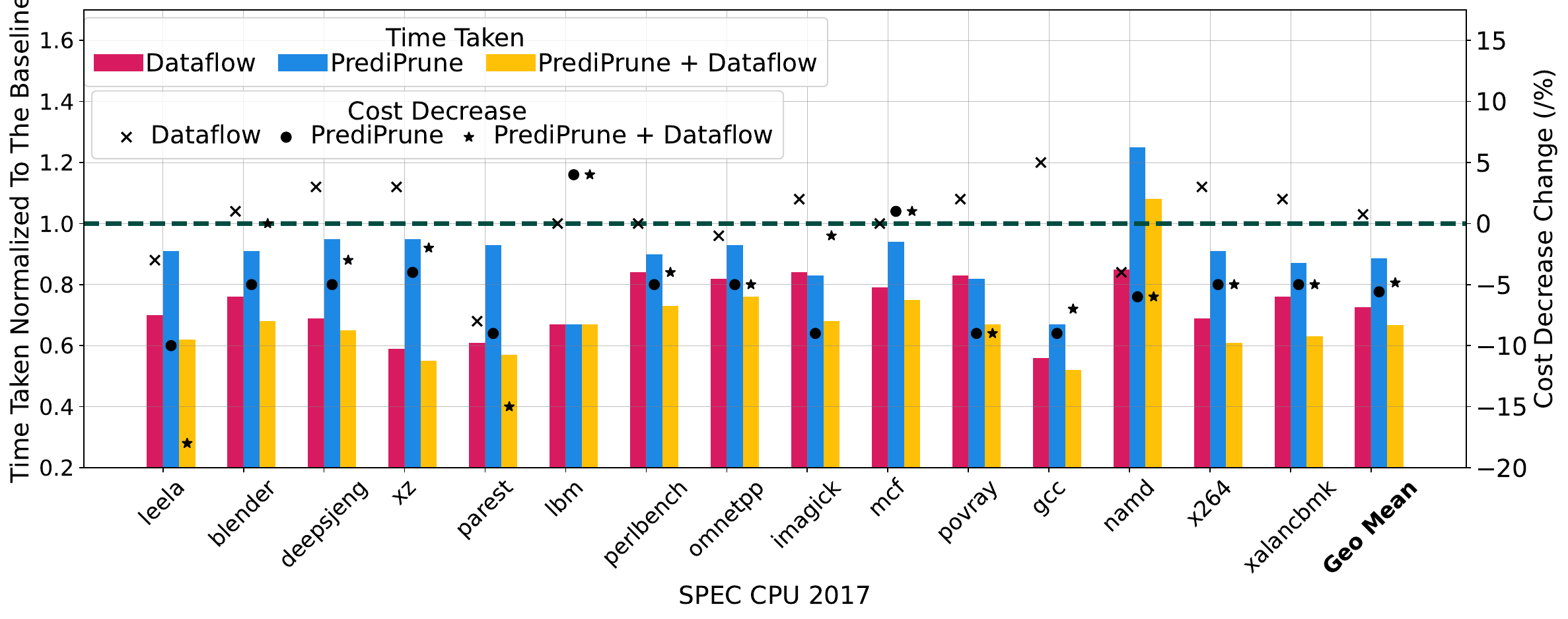}
        \caption{Results of running the SPEC CPU workloads (only C and C++) with the external cache configuration.}
        \label{fig:evaluationCache}
    \end{subfigure}
\end{figure}

\begin{table}
    \centering 
    \begin{tabular}{ |c|c|c|c|c| }

     \cline{2-5}
     \multicolumn{1}{c|}{} & 
        \makecell[c]{\textbf{Time} \\ \textbf{(hrs)}} & 
        \makecell[c]{\textbf{Cost} \\ \textbf{Decrease}} & 
        \makecell[c]{\textbf{Total} \\ \textbf{Success}} & 
        \makecell[c]{\textbf{Time} \\ \textbf{out}} \\
     \hline
     
     Baseline   & 654 & 38\% & 1042 & 6897 \\ \hline
     \rowcolor{lightgray} Baseline with cache  & 384 & 40\% & 1156 & 4463 \\ \hline
     Dataflow & 363  & 42\% & 1087 & 3440 \\  \hline
     \rowcolor{lightgray} Dataflow with cache  & 273  & 41\% & 1094 & 3167 \\ \hline
     PrediPrune & 627  & 37\% & 948 & 6734 \\ \hline
     \rowcolor{lightgray} PrediPrune with cache & 334 & 35\% & 1048 & 3188 \\ \hline
     PrediPrune$+$Dataflow & 320  & 42\% & 963 & 3145 \\ \hline 
     \rowcolor{lightgray} PrediPrune$+$Dataflow with cache  & 244  & 36\% & 1062 & 2908 \\ \hline
    
    \end{tabular}
    \caption{Summary of experimental results with (highlighted in gray) and without the external cache.}
    \label{tab:results}
\end{table}

\begin{table}
    \centering 
    \begin{tabular}{ |c|c| }

    \hline
    \textbf{Configuration} & \textbf{Geo Mean Pruning Rate} \\ 
    \hline
    PrediPrune              & 42\% \\ 
    \rowcolor{lightgray} PrediPrune with cache   & 71\% \\ 
    \hline
    PrediPrune + Dataflow              & 50\% \\ 
    \rowcolor{lightgray} PrediPrune + Dataflow with cache    & 41\% \\ 
    \hline

    \end{tabular}
    \caption{Summary of the pruning rates achieved by PrediPrune.}
    \label{tab:pruning_results}
\end{table}

We experiment with \emph{Souper} without the external cache, to understand the isolated impact of the pruning process. Without the cache, \emph{Souper} must generate and validate all optimization candidates from scratch.
Without the cache (shown in Figure~\ref{fig:evaluation}, and averages in Table~\ref{tab:results}), 
\emph{PrediPrune $+$ Dataflow} achieves a significant reduction in compilation time compared to both \emph{Dataflow} and the \emph{Baseline} methods.
The combined approach reduces compilation time by 51\% compared to the \emph{Baseline} (from 654  hours down to 320 hours to compile the entire SPEC CPU 2017 benchmark suite) and by 12\% compared to \emph{Dataflow} (from 363 hours down to 320 hours to compile the entire SPEC CPU 2017 benchmark suite). 
This demonstrates the effectiveness of \emph{PrediPrune} in eliminating invalid candidates that \emph{Dataflow} cannot, thereby further reducing compilation time by narrowing the verification space. 

While \emph{PrediPrune $+$ Dataflow} improves compilation speed, it maintains the same level of cost reduction.
Both \emph{PrediPrune $+$ Dataflow} and \emph{Dataflow} achieve a 42\% cost decrease compared to the baseline’s 38\% (see Figure~\ref{fig:evaluation}, where the right axis shows the additional cost decrease change based on the Baseline (Souper) cost improvements), highlighting the effectiveness of \emph{PrediPrune $+$ Dataflow} in pruning more invalid candidates while retaining valid ones.

However, we observe that in certain cases -- such as with the \emph{parest} and \emph{leela} benchmarks -- \emph{Dataflow}, \THISWORK{}, and \emph{PrediPrune $+$ Dataflow} lose a portion of the cost savings achieved by Souper (approximately 12\% on \emph{parest} and 7\% on \emph{leela}). One contributing factor is that the Baseline implementation is able to optimize 6 additional LHS in \emph{parest} and 12 in \emph{leela}. In contrast, the pruning step in \THISWORK{} eliminates these opportunities by predicting them to be invalid, while both \emph{PrediPrune $+$ Dataflow} and \emph{Dataflow} encounter timeouts that prevent these additional optimizations.

The significant compilation time improvements in the combined approach of \emph{PrediPrune $+$ Dataflow} is attributed to the additional pruning provided by PrediPrune as shown in Table~\ref{tab:pruning_results}. In 
\emph{PrediPrune $+$ Dataflow}, PrediPrune is able to prune an additional 50\% of the RHS candidates left by \emph{Dataflow}.
By effectively pruning most of the invalid candidates, \emph{PrediPrune $+$ Dataflow} significantly reduces the number of RHS candidates that the SMT solver needs to process.

In contrast to the combined solution, \emph{PrediPrune}, when used alone, does not result in a significant reduction in compilation time.
The pruning rate achieved by \emph{PrediPrune} in isolation is 42\%, which is lower compared to \emph{PrediPrune $+$ Dataflow}'s rate of 50\%. 
This lower pruning rate translates to minimal gains in compilation time.
Additionally, the smaller cost reduction suggests 
that \emph{PrediPrune} inadvertently prunes valid candidates. 
This is further supported by the overlap in the PCA visualization of valid and invalid candidates in Fig.~\ref{fig:selectK}, which highlights the challenges of classifying valid and invalid candidates without applying preliminary pruning. 
However, pruning deterministic heuristic solutions appears to improve the performance of \emph{PrediPrune}, making it more effective when combined with \emph{Dataflow}.

%% file: text/720_evaluation_cache.tex
\subsection{Performance with an External Cache}\label{720_evaluation_cache}

We run the experiments with the external cache connected to Souper.
The external cache significantly reduces the compilation time across all four methods—\emph{Baseline}, \emph{Dataflow}, \emph{PrediPrune}, and \emph{PrediPrune $+$ Dataflow};  the external cache saves time by storing previously generated RHS candidates and their outcome along with their LHS,
so when the same LHS and their RHS candidates re-appear, they can be retrieved from the cache instead of being recomputed.

The external cache significantly reduces the compilation time, as less number of candidates have to go through the SMT solver (results shown in Figure \ref{fig:evaluationCache} and Table \ref{tab:results}). 
The compilation time is reduced by 41\% for the \emph{Baseline} method (from 654 hours to 384 hours), 25\% for \emph{Dataflow} (from 363 hours to 273 hours), 47\% for \emph{PrediPrune} (from 627 hours to 334 hours) and 24\% for \emph{PrediPrune $+$ Dataflow} (from 320 hours to 244 hours), as shown in Table \ref{tab:results}.

The combined \emph{Dataflow} and \emph{PrediPrune} approach still outperforms both \emph{Dataflow} and the \emph{Baseline} when using the external cache, in terms of faster compilation times. 
Specifically, as shown in Table \ref{tab:results}, \emph{PrediPrune $+$ Dataflow} achieves a 36\% faster compilation time compared to the \emph{Baseline} (from 384 hours to 244 hours) and an 11\% faster compilation time compared to \emph{Dataflow} (from 273  hours to 244 hours).
The faster compilation time is attributed to \THISWORK{} pruning 40-50\% of the remaining candidates left by \emph{Dataflow} in the combined configuration, as shown in Table \ref{tab:pruning_results}.
These results demonstrate that \emph{PrediPrune $+$ Dataflow} remains the most efficient approach, even with the external cache in place.

However, not all benchmarks in \emph{PrediPrune} and \emph{PrediPrune $+$ Dataflow} experience significant benefits compared to the \emph{Baseline}, particularly \emph{namd}. A closer analysis reveals that \emph{namd} is the only benchmark where 
substantial compilation time is spent on LHS candidates that ultimately hit the time limit, increasing overall compilation time without yielding meaningful benefits. In comparison, \emph{PrediPrune} and \emph{PrediPrune $+$ Dataflow} encounter 18 and 5 additional LHS candidates, respectively, compared to the Baseline, resulting in an extra 1.5 hours for \emph{PrediPrune} and 0.4 hours for \emph{PrediPrune $+$ Dataflow} in compilation time.
These time limits are first encountered during the run without the cache. Although the first time limits are encountered in the run without the cache, the subsequent run with the cache skips all previously inspected RHS candidates before the time limit and resumes processing the RHS candidate that initially hit the time limit. If this RHS cannot be proven to be semantically equivalent within the time limit while also pruning valid candidates, the LHS will not result in an optimization.
Lastly, both \THISWORK{} and \emph{PrediPrune $+$ Dataflow} experience a drop in the
cost decrease of roughly 5\% compared to the Baseline since they are able to optimize fewer LHS and in addition the majority of the saved cost is trivial compared to the gains made by \emph{Baseline} and \emph{Dataflow}.
Additionally, \emph{parest} and \emph{leela} continue to prune valid candidates, leading to missed opportunities to reduce the cost as seen and addressed in the previous section.

%% file: text/800_related.tex
\section{Related Works}\label{800_related}

Superoptimization seeks to discover the most efficient code sequences that are semantically equivalent to given program fragments. 
Managing the vast search space of potential candidate programs is a significant challenge, necessitating effective pruning strategies to make the process computationally feasible. 
Various pruning techniques have been proposed over the years, each with its advantages and limitations. This section reviews different pruning strategies and explains how {\THISWORK} advances the field by addressing the shortcomings of previous approaches.

\paragraph{Test case-based pruning}
The earliest superoptimizers, such as Massalin's work in 1987~\cite{massalin1987superoptimizer}, relied on a primitive pruning strategy that tested synthesized candidate programs on a set of inputs.
Candidates that did not produce correct results, as verified by comparing outputs with the original program (serving as an oracle), were discarded. 
While this method was simple and could quickly eliminate obviously incorrect candidates, it had significant limitations. 
Testing against a finite set of inputs could not guarantee semantic equivalence, leading to potential false positives. 
Moreover, this approach was not scalable and was unsuitable for compilers and safety-critical applications, where correctness over all inputs is mandatory. 
In contrast, {\THISWORK} avoids reliance on test cases by employing a machine learning model that predicts the validity of candidates based on extracted features, ensuring broader coverage and reducing the risk of false positives.

\paragraph{Solver-based pruning}
To overcome the limitations of test case-based methods, subsequent approaches widely adopted Boolean Satisfiability (SAT)~\cite{solvers2021conflict} and Satisfiability Modulo Theories (SMT) solvers~\cite{schkufza2013stochastic, churchill2017sound, jangda2017unbounded} for equivalence checking. 
These solver-based methods provided formal guarantees of semantic equivalence across all inputs and were suitable for verifying complex program transformations. 
However, they introduced high computational costs because each candidate required a solver invocation, which was time-consuming, especially with a large number of candidates. 
Solvers often processed candidates sequentially, leading to increased compilation times. 
{\THISWORK} addresses these limitations by reducing the number of solver calls. 
By preemptively pruning unlikely candidates using an ML-based model, it decreases the verification workload and speeds up the compilation process.

\paragraph{Speeding up the search process}
GreenThumb~\cite{phothilimthana2016greenthumb} investigated various optimization techniques, including enumeration, stochastic, and symbolic synthesis, to enhance the search process and more efficiently generate
candidate optimizations. 
While these methods combined multiple synthesis strategies to explore the candidate space more effectively and had the potential to find optimizations that single-method approaches might miss, the validation step remained a bottleneck. 
The high number of candidates still required expensive solver checks, and mixed techniques could generate more candidates, increasing the likelihood of invalid ones slipping through initial pruning. 
{\THISWORK} addresses this validation bottleneck by significantly reducing the number of candidates needing solver validation. 
Its ML-based pruning filters out unlikely candidates early, streamlining the entire optimization pipeline.

\paragraph{Sketching and metasketching}
Some researchers proposed sketching and metasketching techniques to reduce the search space by dividing it into multiple smaller, ordered sketches and employing parallelized search algorithms~\cite{bornholt2016optimizing}, again resulting in faster candidate generation. 
While this approach exploited parallel computation to accelerate the search process and organized the search space to potentially find optimizations faster, it introduced additional computational overhead in managing multiple sketches and resolving local optima. 
Implementing such techniques required sophisticated mechanisms to divide and coordinate search spaces effectively. 
In contrast, {\THISWORK} simplifies the approach by using machine learning to guide pruning without the need for complex search space division. This results in less overhead and a more straightforward integration into existing compiler infrastructures.

\paragraph{Dataflow analysis-based pruning}

Dataflow analysis-based pruning techniques emerged as a way to manage the candidate space more efficiently, and is the most directly comparable strategy to {\THISWORK}. 
The data pruning approach introduced in Phothilimthana et al. ~\cite{phothilimthana2016scaling} used bidirectional dataflow analysis to prune the candidate space, achieving up to $11\times$ speedup over solver-only solutions. 
Mukherjee et al.~\cite{mukherjee2020dataflow} extended this approach to \emph{Souper}. 
While these methods were more efficient and scalable compared to solver-only methods, they relied on deterministic heuristics that might not generalize well across different code patterns. 
They also had limited pruning coverage and might not effectively prune all invalid candidates, leading to unnecessary solver calls. 
{\THISWORK} builds upon \emph{Souper} but employs stochastic methods via machine learning to achieve broader pruning coverage. 
By extracting type-agnostic features and using an MLP classifier, it effectively reduces verification overhead while maintaining or improving optimization opportunities.

\paragraph{Machine learning driven optimizations}
Machine learning has also proven useful for optimizing various parts of the compilation process~\cite{mendis2019compiler,mendis2019ithemal, venkatakeerthy2024next, mithul2024exploring, chen2021efficient}. Ithemal~\cite{mendis2019ithemal} 
is a tool that predicts the number of cycles a basic block will take to execute based on its opcodes and operands, helping engineers write more efficient code. Another example is BOCA~\cite{chen2021efficient}, which leverages a Random Forest model to efficiently identify the most effective compiler flag combinations.

%% file: text/900_conclusion.tex
\section{Conclusion}\label{900_conclusion}

{\THISWORK} advances the field of superoptimization by introducing a machine learning-based pruning strategy that overcomes the limitations of previous approaches. 
By leveraging stochastic methods and type-agnostic features, it achieves broader pruning coverage and significantly reduces verification overhead. 
Its approach provides a flexible trade-off between compilation time and optimization effectiveness, making it superior to traditional deterministic heuristics and more adaptable for integration into existing compiler infrastructures.